\documentstyle[11pt]{article}
\newcommand{\pbar}{\overline{p}}
\newcommand{\pbarp}{\overline{p}p}
\newcommand{\nbarn}{\overline{n} n}
\newcommand{\pbarpnbarn}{\overline{p}p \rightarrow \overline{n}n}
\newcommand{\NbarN}{\overline{N}N}
\newcommand{\qbarq}{\overline{q}q}
\begin{document}
\begin{center}

\title{What have we learned from antiproton proton scattering?}

\vspace{8mm}

\author{
F. Myhrer\thanks{This work is supported in parts by the National Science
Foundation, Grant No. PHYS-9310124}
 \\
Department of Physics and Astronomy, \\
University of South Carolina, \\ 
Columbia, SC 29208, USA }

 
\maketitle

\vspace{30mm}

{\bf Abstract}
\end{center} 
 From recent charge exchange measurements in the extreme 
forward direction, 
an independent and precise determination of the pion
nucleon coupling constant is possible. 
This determination has reopened the debate on the value of this 
fundamental coupling constant of nuclear physics.
Precise measurements of 
charge exchange observables {\it at forward angles} 
below 900 MeV/c would also give a better understanding  
of the long range part of the two-pion exchange potential. For example, 
the  confirmation of the coherence of the tensor forces from the pion 
exchange and the isovector two-pion exchange would be very valuable. 
With the present data first attempts at an $\NbarN$ partial 
wave analysis have been made 
where, as in nucleon nucleon scattering, the  antinucleon nucleon 
high $J$ partial waves are 
mainly given by one-pion exchange. 
Finally a recent $\pbarp$ atomic cascade calculation and 
the fraction of P-state annihilation in gas targets is commented on.

hep-ph/9611381

\newpage

\section{INTRODUCTION}
In this presentation which is dedicated to the memory of Carl Dover, 
I shall discuss what low energy antiproton proton scattering data 
have contributed or can 
contribute to our knowledge of hadronic physics. 

\vspace{3mm}

The low energy strong interactions are governed by chiral symmetry,
an approximate but very good symmetry  of the QCD lagrangian.
This symmetry is perfect if the u- and d- quarks are massless. In reality
the quark masses are not zero but since $m_u \approx m_d \ll \Lambda_{QCD}$
or since the pion mass $m_{\pi} \ll \Lambda_{\chi}$, 
where the chiral symmetry scale $\Lambda_{\chi} 
\sim $ 1 GeV, we expect this symmetry to be very good.

The successes of chiral symmetry (current algebra) 
predictions when confronted with the experimental
measurements have been very successful.
The next order 
corrections to these lowest order predictions as 
calculated in chiral perturbation 
theory (ChPT), indicate that we have 
achieved a much deeper understanding of how chiral symmetry is 
broken in hadronic phenomena. 
Simultaneously the evolution of the 
chiral quark models, which implements chiral symmetry 
on the quark level in hadronic models, leaves us with the 
following model understanding of the structure of the nucleon: 
The nucleon consists of quarks confined to a core of radius 
of the order 0.8 fm, a core which by necessity 
(due to chiral symmetry) is 
surrounded by a pionic (Goldstone boson) cloud. 
%
This pionic cloud is distributed around the quark core according to the
requirements of the (approximate) chiral symmetry.

This ``chiral" nucleon model predicts that 
for large impact parameter scattering of 
two nucleons ($NN$) or antinucleon-nucleon ($\NbarN$), 
only the pionic clouds will overlap. 
Consequently, we expect that for these large impact parameter scatterings 
the effective meson exchange potential models should give a reasonable 
description of the scattering data and 
we can relate the
$NN$ and $\NbarN$ meson exchanges using G-parity invariance arguments.
 From these arguments we find 
that the one-pion-exchange (OPE) potential, $V_\pi$, has opposite
signs for $NN$ and $\NbarN$, whereas the two-pion-exchange (TPE) 
potential, $V_{2 \pi}$, has the same sign for both $NN$ and $\NbarN$. 
In the following we shall discuss the consequences and some experimental 
confirmations of these ideas. 

\section{THE STRENGTH OF THE PION NUCLEON COUPLING} 
{\it The pion nucleon coupling constant is
a fundamental constant in nuclear physics. 
It determines the strength of the very important
long distance nuclear tensor potential.}

The recent very precise 
charge exchange measurements ($\pbarpnbarn$) of Birsa et al. \cite{Birsa94,amartin96,Bressan}
in the extreme forward direction has established 
beyond any doubt the
dominance of the one pion exchange between the antinucleon and the nucleon. 
These data were presented at this conference by A. Martin \cite{amartin96}. 
As expected, the extreme forward scattering cross section is dominated by 
the OPE pole which, at four momentum transfer $t$ = + $m^2_{\pi}$, 
is located just below the physical region. 
By extrapolating these data at 
forward scattering angles ($0 < -t < 1.5  m^2_{\pi}$) to the pion pole, 
the residue of the pole is evaluated, 
i.e., the pion-nucleon coupling constant,
$g_{\pi NN}$, is determined. 
The assumption here is that 
\begin{equation}
(m_{\pi}^2 - t)^2 \cdot \frac{{\rm d}\sigma}{{\rm d} \Omega} 
= \frac{1}{E_{cm}^2}\sum_{n=o}^\infty a_n (m_{\pi}^2 - t)^n , 
\end{equation}
is a smooth polynomial in $(m_{\pi}^2 - t)$ and can be extrapolated to
$t = m_{\pi}^2$ \cite{amartin96}. 
In the above expression $E_{cm}$ is the center of mass energy.
In fact, these 
new data give a very precise value for $g_{\pi NN}^2/4 \pi$
\cite{Birsa94,Panic96a,Franco95},
a value which is smaller  than the
accepted value by about 5\%.
Recently the Nijmegen group, who analyzed both the $NN$ data 
\cite{NijmegenN} and the $\NbarN$ data \cite{Nijmegen} 
within a OBEP model, 
also advocated a smaller value for $g_{\pi NN}^2/4 \pi$. 
Due to this development, the
nuclear physics community is 
re-examining the previous determinations of $g_{\pi NN}$,  
by repeating some of the experiments relevant for its determination as 
discussed below.

The pion nucleon coupling constant has been \cite{Uppsala}, 
and will be \cite{Panic96b} determined from
the extreme forward angles of the 
reaction $ n p \rightarrow p n$ 
which like $\pbarpnbarn$ is dominated by the OPE pole just beyond the 
physical region.
The Uppsala group redid the 
 $ n p \rightarrow p n$  experiment. 
They compared various $n p \rightarrow p n$ experiments and 
showed that several experiments are 
not as accurate as was originally claimed \cite{Uppsala,Panic96b}, 
the normalization (systematic) errors 
of some experiments
are more uncertain than what was quoted in the publications. 
New experiments paying particular attention to the normalization errors are
presently being considered/performed/analyzed at 
several laboratories.
We look forward to the results of these 
experiments and analysis. 

The coupling constant $g_{\pi NN}$ is also determined from the 
$\pi^{\pm} p$ total cross section data 
using dispersion theory (and analyticity). 
For $\pi N$ scattering Locher and Sainio
in a contribution to the PANIC'93 conference 
\cite{PANIC93} discussed the possible
uncertainties coming from the integral part of the 
dispersion relation expression.
They conclude that the dispersion integral itself 
contributes a value with very small
uncertainty to $g_{\pi NN}$. 
The isovector $\pi N$ scattering length, which enters 
this expression for $g_{\pi NN}$, is also very well determined. 
We should keep in mind that 
the value of $g_{\pi NN}$ determines the strength of 
the long range nuclear potential ($V_\pi$) which gives, for example, 
the dominant contribution to the measured 
deuteron asymptotic D/S ratio \cite{WeisEbook}.

These three independent reactions should of course 
give the same value for this coupling constant. 
Presently the $\pbarpnbarn$ appears to give a
smaller coupling constant ($g_{\pi NN}^2/4 \pi \approx $ 13.7) 
about 5\% smaller  than the other two 
methods ($g_{\pi NN}^2/4 \pi \approx $ 14.4). 
This will be discussed
more in detail in a contribution by T. Ericson \cite{Ericson96} 
to this conference.
It should be remarked that at this level of precision (accuracy) 
the Coulomb corrections of the $\pbarpnbarn$ reaction 
should be carefully considered. 
In addition, for the accuracy of $g_{\pi NN}$ being discussed, 
we possibly should discuss the difference of  
$g_{\pi^+pn}$ and $g_{\pi^o pp}$ as well.

An experimentally 
precisely determined  value of $g_{\pi NN}$ is necessary 
in order to judge the accuracy of the theoretical calculated 
$g_{\pi NN}$ value using ChPT. 
Theoretically we know that $g_{\pi NN}$ is given to lowest
order (when $m_u = m_d = m_{\pi}$ = 0) by the Goldberger Treiman relation
\begin{equation}
g_{\pi NN}  = \frac{g_A}{f_{\pi}} \cdot M \approx 12.7 ,
\end{equation}
a value which is smaller than the measured value.
Here the nucleon axial coupling constant, $g_A$, the 
pion decay constant, $f_\pi$, and the nucleon mass, $M$, are 
the physical constants in the ChPT effective lagrangian. 
However, 
chiral symmetry is not a perfect symmetry and the loop corrections to 
the r.h.s. of this relation will increase the value of $g_{\pi NN}$. 
The corrections due to the broken chiral symmetry may be evaluated
in ChPT, but do depend on low energy constants, see e.g. 
Bernard et al. and others \cite{Bernard}. 

\section{THE MESON-EXCHANGE MODELS}
{\it What more can be learned from $\NbarN$ scattering 
relevant to nuclear physics ?}

\subsection{The two-pion exchange potential}
The charge exchange reaction is a unique testing ground for the 
meson exchange potential (MEP) 
models since this reaction is given by the differences of two large
isospin amplitudes. 
The long range part of $V_{2 \pi}$ is not well known, but 
we know that $V_{2 \pi}$ has a range of $(2 m_\pi)^{-1}$ or shorter 
as is evident from the following expression: 
\begin{equation}
V_{2\pi}^J \sim \int_{2 m_{\pi}}^\infty  \frac{e^{-\mu r}}{r} 
\rho^J(\mu) d\mu  , 
\end{equation} 
where the mass spectral function, $\rho^J(\mu)$, 
includes the $J$ = 0 or 1 $\NbarN \rightarrow \pi \pi$ 
helicity amplitudes.  
Theoretically the 
two-pion exchange potential 
has been calculated from $\pi N$  and $ \pi \pi$ scattering data using 
dispersion theory and analyticity to determine $\rho^J(\mu)$ 
in Eq.(3) \cite{Stonybrook,Paris} 
or has been constructed from  effective meson-baryon models \cite{Bonn}. 
(As remarked at the conference: 
using dispersion theory and the $NN$ total cross section data 
in pure $NN$ spin states \cite{GreinK80}, 
the $2 \pi$ $NN$ exchange 
contribution can be extracted and thereby used 
to evaluate $V_{2 \pi}$, 
or more precisely to test $\rho^J(\mu)$ in Eq.(3).) 
These theoretically determined 
long range parts ($ < (2m_{\pi})^{-1}$) of the
correlated isoscalar two-pion S-wave and isovector two-pion P-wave, 
predict the $\NbarN$ observables at forward angles,
and these ``predictions" can now be confronted with 
$\NbarN$ forward angle scattering experiments.

The differential cross section for the charge exchange reaction
has a characteristic dip-bump (minimum -  second maximum) 
structure at forward angles as now firmly established by the 
LEAR experiment, PS206, of Birsa et al. \cite{Birsa94}. 
As discussed by 
Phillips \cite{Phillips} this behavior of the cross section 
with a second maximum ``is a typical OPE effect, coming from 
the double spin-flip amplitude". 
However, as will be clear from the 
arguments below, we can extract information about $V_{2 \pi}$ 
from this particular angular behavior of the differential cross 
section for $\pbarpnbarn$.  

In terms of the five helicity amplitudes the differential cross section 
for the charge exchange reaction is 
\begin{equation}
\frac{{\rm d}\sigma}{{\rm d}\Omega} = | \phi^{cex}_1|^2 + 
|\phi^{cex}_2|^2 + |\phi^{cex}_3|^2 + |\phi^{cex}_4|^2 + |\phi^{cex}_5|^2 
\end{equation} 
In the Born approximation for most one-boson-exchange models 
the double helicity flip amplitudes $\phi^{cex}_2$ and $\phi^{cex}_4$
are dominated by OPE.  
In Eq.(4) the amplitude $\phi^{cex}_2$ has a strong forward peak 
whereas $\phi^{cex}_4$ is zero
at zero degrees and increases to a maximum at larger angles 
(30$^{\rm o}$-60$^{\rm o}$). 
The sum of these two OPE dominating helicity amplitudes generates the 
dip-bump structure of the forward differential cross section.
The point is that $V_{2 \pi}$ (which is a sum of an isoscalar, 
$V_{2 \pi}^0$, and an isovector, $V_{2 \pi}^1$, potential) 
also contribute to $\phi^{cex}_2$ and $\phi^{cex}_4$. 
However, as stated, in the Born approximation 
both amplitudes are dominated 
by OPE but with a smaller and important contribution from 
the shorter range ``$\rho$-meson" 
exchange or more precisely 
the TPE isovector part, $V_{2\pi}^1$, of Eq.(3) \cite{MMT85}.  
This potential  
includes in addition to the ``bare" ($\overline{q} q$) $\rho$-meson 
exchange, the long range correlated isovector TPE which is 
described by $\rho^J(\mu)$ of Eq.(3). 
In other words, $V_{2\pi}^1$ describes a 
"bare" $\qbarq$ $\rho$-meson
embedded in the correlated isovector two-pion continuum \cite{Stonybrook}. 
At the forward angles ($< 40^{\rm o}$ at p$_{lab}$ = 600 MeV/c) 
under discussion
the three other helicity amplitudes in Eq.(4) are 
given by ``$\rho$-exchange" in the Born approximation. 
These three amplitudes  are smaller by a factor 2-10 depending on 
the MEP model used in the calculation of these amplitudes  \cite{MMT85}. 
Given the accuracy of the PS206 data \cite{Birsa94}, it 
is the exploration of this minimum, typical of OPE but modified by the 
long range part of TPE where we can extract more information about 
$V_{2\pi}$ itself. 

As we shall discuss next, the 
effects of $V_{2 \pi}$ also manifest themselves in other 
$\NbarN$ observables
which will contain further information about 
the long distance part of $V_{2 \pi}$.
Two other observables, $A_{on}$ and $D_{onon}$, can give valuable 
information about $V_{2 \pi}$ and complement the information from 
${\rm d}\sigma/{\rm d}\Omega$ just presented.
Both $NN$ and $\NbarN$ have an intermediate range attraction 
due to the two-pion isoscalar exchange potential, $V^0_{2 \pi}$.
The strength of the isoscalar 
two-pion exchange at intermediate 
distances has an important $\vec{L} \cdot \vec{S}$ component 
which, together with the strong, coherent $\NbarN$ $I$ = 0 
tensor part of the $V_{\pi}$ and $V^1_{2 \pi}$, 
accounts for the very rapid rise from zero of the
measured analyzing powers, $A_{0n}$, in $\pbarp$ elastic, 
see e.g. Ref.\cite{Bertini}, and charge exchange
reactions at small forward angles. 
(For $NN$ the two tensor potentials, from $V_\pi$ and from $V^1_{2\pi}$,  
have opposite signs and cancel at medium $NN$ distances.)
In addition,  
as can be clearly seen in Ref.\cite{thesis94},e.g., by comparing 
Figs. 10.1 through 10.6 of Ahmidouch's Ph.D. Thesis \cite{thesis94}, 
all MEP models predict a 
maximum peak at non-zero but forward angles in the $D_{onon}$ 
$\pbarpnbarn$ observable. 
This forward $D_{onon}$ peak is again mainly due to 
$V_{\pi}$ and $V_{2\pi}$. 
In other words {\it for
both $A_{on}$ and $D_{onon}$
$\NbarN$ observables, meson exchanges are 
clearly seen at forward angles.}
(In OBEP models the continuous mass distribution of $V_{2\pi}$ is 
simulated by $\sigma$ and $\rho$
exchange potentials, e.g. \cite{Nijmegen}.)
These two features of the 
two $\pbarpnbarn$ spin observables, 
the rapid increase of $A_{on}$ for increasing scattering angles 
and the peak at forward angles of $D_{onon}$,  
would be reasonable experimental testing grounds for the expectation that 
the isovector TPE tensor potential adds coherently to the 
tensor part of OPE, 
as stressed by Carl Dover and Jean-Marc Richard \cite{Dover82}. 
We do, however, need to measure $D_{onon}$ at forward angles 
($< 30^{\rm o}$ for p$_{lab} <$ 900 MeV/c) 
to determine the maximum value of the ``predicted" peak.  

In short the $\pbarpnbarn$ reaction is very promising 
to accurately determine both $V_{\pi}$ and $V_{2\pi}$ 
and to test the tensor part of $V_\pi$ plus ``$\rho$-exchange".
To experimentally test the theoretically long range predictions of 
$V_{2 \pi}$, and to separate better the $V_\pi$ effects, 
we need to measure, a la PS206, the
energy dependence
of the forward ${\rm d}\sigma /{\rm d}\Omega (\pbarpnbarn)$ 
minimum, we need measurements 
for at least two more momenta in the range $p_{lab}$
$\sim$ 200 - 800 MeV/c. 
%
%
{\it 
Here the lower LEAR energies 
are emphasized since the concept of the OPE- and TPE potentials 
and their connections to nuclear matter 
are only reasonable for not too large energies and momentum 
transfers.} 

The implication of a better known $V_{2 \pi}$ is as follows.
In $NN$ models the $V_{2\pi}$ attraction must always be ``balanced" 
by models for the unknown 
short range $NN$ repulsion, a repulsion which 
contributes to the stiffness of the equation of state
of nuclear matter. 
(The point of ``balancing" the intermediate $V_{2\pi}$ attraction 
and the short range repulsion in $NN$, 
is most easily seen in OBEP models: 
There is a close relation
between the strong short range repulsion of $\omega$ exchange 
and intermediate attraction of the scalar isoscalar ($\sigma$) exchange,
which simulates the scalar two-pion exchange. For very low
three-momentum transfer $ |\vec{q} | \ll m_{\omega}$ or 
$ m_{\sigma}$, the OBEP
coupling constants of $\omega$ and $\sigma$ exchange have 
to be related by 
$g_{\omega} ^2/m_{\omega}^2 = g_{\sigma}^2/ m_{\sigma}^2$ so that
only $V_\pi$ effects remain at large distances.) 
Therefore, a  better known $V_{2\pi}$ would indirectly 
give information about 
how stiff is the nuclear equation of state. This would allow us to better
predict what happens with nuclear matter at high densities 
which is believed to occur during, for example, supernova explosions.

\subsection{The short range $\NbarN$ phenomena.}
{\it Do the short 
range annihilation reactions modify the previous discussions?} 

Based on our model understanding of the nucleon discussed in the introduction,
we expect
annihilation to occur only for low
impact parameter scattering 
when the quark cores of $N$ and $\overline{N}$ overlap \cite{Amsler}. 
In other words, we expect $\NbarN$ annihilation to have a much shorter range 
than OPE. However, the 
$\rho$ meson has not only 
a large width, the ``$\rho$-exchange" has 
a range as large as $(2 m_\pi)^{-1}$, as
is evident from Eq.(3), thereby 
creating a long range $\NbarN$ $V_{2\pi}$ tail 
which extends beyond the annihilation region. 
By contrast, the $\omega$ meson has a very small width. 
Therefore, a short range meson exchange, 
like $\omega$ exchange, becomes masked by the short
range but violent annihilation reactions. 
However, in $\pbarpnbarn$ observables like $A_{on}$ we
may see traces of $\omega$ exchange 
provided the effective model coupling constant 
$g_\omega^2/4 \pi$ is forced to be 
very large $>10-20$ \cite{Mull}. 
Below we shall present the ideas behind the various annihilation 
models. 

Phenomenologically we know that these model ideas just presented, 
of a nucleon consisting of a quark core surrounded by a pion cloud, 
are reasonable. 
This is based on, for example, the
quark model calculations of Oka and Yazaki and others, who  
 successfully reproduce
the energy behavior of the $NN$ S-wave phase-shifts, for 
reviews see, e.g. Refs.\cite{NNquarks}. For $NN$ S-wave 
(zero impact parameter) 
scattering the two nucleons feel the meson exchange forces
at large separations. But at shorter $NN$ distances, when the 
two quark cores overlap, 
the antisymmetry requirement of the overlapping six quark
 wave functions
(with a little help from one gluon exchange to get the correct
 N- $\Delta$ mass difference) generates 
the observed hard-core-like S-wave phase shifts 
for increasing $NN$ energy.
Our chiral quark model, which naturally includes meson exchanges, 
has a short distance $NN$ repulsion which
arises from quark antisymmetry requirements, and 
which easily accommodates $\omega$ exchange with 
the standard SU(3) value for
$g_{\omega}^2/4 \pi$ = 4.5. 
In most MEP models, which are used in $\NbarN$ calculations, 
a value  $g_{\omega} ^2/ 4 \pi$ = 10 - 12 is required when 
the $NN$ short range repulsion is
ascribed to $\omega$-exchange alone. 
These models \cite{Paris,Bonn,Nij75} necessarily include form factors
and/or short distance parametrizations, 
but as discussed above modify to some extent 
the effects of $V_\pi$ and $V_{2 \pi}$.

In $\NbarN$ scattering only the bulk
properties of the apparently dominating annihilation reactions 
are of importance. 
In a Skyrme model inspired calculation of $\NbarN$ annihilation,  
the bulk properties of the many pions produced have been 
reasonably reproduced, but 
annihilation appears to take place as soon as the 
$N$ (soliton, size of about 1 fm) and $\overline{N}$ (antisoliton) touch, 
see Ref.\cite{Amado} and references therein. 
Within our chiral quark model we can understand this 
seemingly large range annihilation by the following example: 
In the simple absorptive ``black sphere" model of radius 0.5 fm, 
the annihilation still appears to have a range of 1.5 fm 
\cite{Dalkarov}. This 
is generated by the long range attractions of  $V_\pi$ and $V_{2 \pi}$ 
which enhance the $\NbarN$ wave function in the 
short distance annihilation region. 

A more practical annihilation description of $\NbarN$ scattering, 
which includes explicitly $V_\pi$ and $V_{2 \pi}$, 
invokes the effective doorway mesons. 
As shown by Vandermeulen \cite{Vanderm}, 
the many specific annihilation 
cross sections have been successfully 
described by postulating that $\NbarN$ first form two doorway
mesons which then decay with known branching ratios into pion and kaon
final states.  He argued that it is necessary to average the spin and isospin
of the intermediate mesons, 
and showed that the pairs of two intermediate mesons 
with sum of masses closest to
the available $\NbarN$ c.m. energy dominate the annihilation process. 
This doorway mesons mechanism can easily account for the 
apparent observed  OZI violations in specific $\overline{p} p$ 
annihilation channels, see e.g. \cite{Locher}. 

Several groups, e.g., \cite{Bonn2,Mouss,Tabakin} are using the doorway mesons 
in coupled channel models where it is imperative that 
``very many mesons" 
channels will contribute to the process such that 
the annihilation processes will contribute only extremely  weak spin and 
isospin dependences to the $\NbarN$ scattering itself.  
Since most of these doorway mesons have widths, 
most meson thresholds will not be reflected in 
sharp cusps like what two stable mesons will generate. 
Also the Nijmegen \cite{Nijmegen} and Moscow \cite{Moscow} 
groups use effective coupled channels 
to include some energy
dependence expected from the annihilation reactions. 
For a short overview of these coupled channel calculations, 
see Ref.\cite{Loiseau}. 
This energy dependence is neglected in the 
standard optical potential 
description of the annihilation processes. 
The common features in these models are (i) the short range of the 
annihilation process itself and 
(ii) the very weak explicit
isospin dependence of the combined $\NbarN$ short range parametrizations and 
annihilation descriptions, a dependence 
which is necessary in order to reproduce the data.

\section{Meson exchange ``model analysis" and $\NbarN$ partial waves.}

Fortunately, we do have some very useful theoretical restrictions 
to start an analysis.
As for $NN$, the very high $\NbarN$ partial wave 
amplitudes are 
dominated by OPE with some TPE corrections. 
The dominance of OPE has now been verified experimentally as discussed. 
Both the Nijmegen \cite{Nijmegen} and the Paris \cite{Parisan} 
groups have made a first attempt at a ``partial wave" analysis of the 
$\NbarN$ scattering data.
These analysis are 
useful in the sense that we learn which aspects of $\NbarN$ scattering data are
sensitive to which parameters of the models used in the analysis. 

There is some controversy regarding the present
$\NbarN$ energy-dependent analysis \cite{debate}.
At a given energy there are not sufficient observables measured yet 
to perform a model independent partial wave analysis.
In the above energy dependent analysis \cite{Nijmegen,Parisan}, 
the models are used (with their 
unknown parameters) to provide the link between measurements at 
different energies. 
To test the present analysis,  measurements of 
${\rm d}\sigma/{\rm d}\Omega$ and $A_{on}$ for p$_{lab} <$ 400 MeV/c 
would be very desirable.

However, we know that different data have different types of 
systematic errors. To ignore data points or whole data sets just
because they give too large $\chi^2$ in a specific model analysis 
is too subjective a criteria and should be avoided. 
We need data with good statistics and with 
reasonable published systematic errors to make a serious 
analysis. As an example of a useful discussion on comparing 
different measurements with different systematic errors, the discussion 
of $n p \rightarrow p n$ by the Uppsala group \cite{Panic96b} should
serve as a model. 
In addition the 
systematic errors should be used to evaluate and re-examine data and 
be included in a $\chi^2$ fit. 
D'Agostini
\cite{errors} gives one indication
of how to include systematic errors in a $\chi^2$ 
analysis of data. This is certainly a less subjective criteria 
to be used in an analysis of published data.

\section{ANTIPROTONIUM CASCADE CALCULATIONS} 

I would like to make a 
short comment on a recent $\pbarp$ atomic cascade calculation and 
the fraction of P-state annihilation in gas targets.
In a recent paper Batty \cite{Batty} compares two different atomic 
cascade model calculations 
and shows how they compare to, e.g. measured P-wave annihilation 
fractions,$f_P$, as a function of the target density, 
and how both atomic cascade models predict an $f_P$ versus 
target density somewhat different from the ``measured" 
$f_P$ especially at low target densities. 

He also discusses the fine structure populations of 
the atomic states and how 
the populations could depend on target density. 
The lowest atomic states 
are not statistically populated due to strong $\pbar p$ interactions. 
However, the lowest $^3P_0$ atomic state is more populated 
at higher target densities 
\cite{Batty}. 
These findings warrant further investigations since they are 
important in order to determine the atomic annihilation channel 
branching ratios into specific final meson states. 
In the $\pbarp$ atom the $\nbarn$ component 
of the atomic wave function
changes rapidly with distance for r $<$ 5 fm. 
Since the annihilation effectively occurs at shorter atomic  $\NbarN$
distances ($\approx$ 1 fm) we need a better understanding 
of the atomic
wave function for r $<$ 5 fm in order to calculate specific relative 
annihilation branching
ratios from specific atomic states ($^{1}S_{0}, ^{3}P_{0}, ^{1}P_{1}$, etc.). 
The precise measurements of $X$-rays from $\pbarp$ and $\pbar d$ 
\cite{Gotta} should contribute towards this understanding.

\section{ CONCLUSIONS}

The new $\NbarN$ scattering data have 
revived the discussion of the $g_{\pi NN}$ value, a fundamental 
coupling constant in nuclear physics.
The value of this coupling constant can 
be evaluated in ChPT. However, the question is if 
the present ChPT correction to Goldberger-Treiman relation, Eq.(2), 
is sufficient, or do we have to consider the next order 
in ChPT? 
We will not know until we know the precise value of $g_{\pi NN}$, 
an accuracy which 
approaches the level where Coulomb and 
isospin breaking interactions might be of importance. 

For the charge exchange reaction 
the dip-bump in the differential cross section 
and the forward angle behavior of 
$A_{on}$ and $D_{onon}$ at
low energies will teach us more about the long range part of 
$V_{2 \pi}$. 
Further precise measurements of both elastic and charge exchange 
observables like $A_{on}$ and $D_{onon}$ {\it at forward angles} 
would contribute to the understanding  
of the long range part of the two-pion exchange. 
As advocated by Carl Dover and Jean-Marc Richard, 
the confirmation of the possible coherence effects 
in the $\pbarp$ $I$ = 0 amplitude 
from the pion exchange and the isovector 
two-pion exchange would be very striking. 
The charge-exchange reaction is 
the most sensitive probe because it is 
given by the difference 
between two large isospin amplitudes. The other $\NbarN$ 
scattering processes are determined by 
either a single isospin amplitude or the sum of two large 
isospin amplitudes. 
As discussed, since 
annihilation dominates the shorter range processes, the heavier
meson exchanges will not be apparent in the scattering data 
at forward angles. 

The meson exchange description presented here 
cannot be valid for large momentum transfer, $\vec{q}$ 
(or small impact parameter scattering),
where not only MEP vertex form factors (quark structure of 
the hadrons) but annihilation processes 
become very important. 
As discussed, in $\NbarN$ scattering prosesses 
only the global properties
of annihilation are of importance but our models for
annihilation are still too primitive to 
generate a reliable physics description of 
$d \sigma / d \Omega$, $A_{0n}$ and $D_{onon}$ 
for large $\vec{q}$ or backward angles when 
$p_{lab} > $ 0.4 GeV/c.

As stated the charge exchange reaction is a unique testing ground 
for the MEP models since this reaction is given by the differences of 
two large isospin amplitudes. Experimentally it is now 
established that for 
the very forward ${\rm d}\sigma/{\rm d}\Omega (\pbarpnbarn)$ 
we have a minimum in the angular distribution, a minimum which 
is due to the interference of two dominant helicity amplitudes 
$\phi^{cex}_2$ and $\phi^{cex}_4$, confirming the inference of 
Phillips \cite{Phillips} that this angular behavior of the forward 
cross section is tied to the long range one-pion-exchange. 
%
This information can now be used to extract more precise 
information on the long range two-pion-exchange 
potential, $V_{2\pi}$,
which is the same in the $NN$ potential. 
However, this part of the $NN$ potential 
has to be balanced with the largely unknown 
strong short range $NN$ repulsion. 
The strength of this short range repulsion is found by 
fitting the $NN$ potential to the measured $NN$ 
phase shifts. This $NN$ potential is then used in 
nuclear matter calculations and the repulsion 
will give the stiffness of the 
nuclear equation of state. A better determined $V_{2\pi}$ 
would therefore indirectly give more reliable knowledge of 
what happens for example in 
the compressed nuclear matter stage of a 
supernova explosion where the stiffness of 
the nuclear equation of state is important. 
In other words the more detailed knowledge of the long range $V_{2\pi}$ is very important in several branches of nuclear physics and 
the $\pbarpnbarn$ reaction is a unique tool to learn more about this 
part of the nuclear forces.

\end{document}